\documentclass[%
 reprint,
 amsmath,amssymb,
 aps,
]{revtex4-2}

\usepackage{graphicx}
\usepackage{dcolumn}
\usepackage{bm}
\usepackage{multirow}
\usepackage{xcolor}
\usepackage{hyperref}
\usepackage{url}
\usepackage{array}
\usepackage{booktabs}

\begin{document}

\preprint{APS/123-QED}

\title{
Field Deviations in Dipole-Driven Linear Paul Traps: Effects of Endcap Boundaries and their Minimization
}
\author{Vaibhav Mahendrakar,$^{1}$ N. Joshi,$^{1}$, S. A. Rangwala$^{1}$}
 
\affiliation{%
 $^1$Raman Research Institute, C. V. Raman Avenue, Sadashivanagar, Bangalore 560080, India\\
 }%
\begin{abstract}
Deviations from both the ideal linear Paul trap (LPT) geometry as well as the ideal quadrupole driving scheme introduce imperfections to ion trapping potentials. We investigate the effects of these imperfections in a LPT operated in a conventional dipole-drive configuration. We demonstrate the trapping of the Li$^+$ ions along the axial direction with zero and negative end-cap voltages. This occurs due to the modified axial $a-q$ space resulting from radial-to-axial coupling of the electric field. The dipole drive configuration lifts the degeneracy of the radial trapping potentials, resulting in unequal radial secular frequencies, and this is demonstrated experimentally. The combined effects of dipole drive and trap dimensions are summarized in a two-dimensional map that quantifies deviations from ideal behaviour. Based on this map, we propose a geometric modification that significantly reduces radial-to-axial coupling of the potential.
\end{abstract}
\maketitle
\section{\label{sec:intro} Introduction}
Linear Paul traps (LPTs) continue to be a foundational tool in atomic physics, with applications spanning mass spectrometry \cite{Douglas,Kitaoka_2013,Hudson}, collisional physics \cite{denschlag2014macro,jian2005macro}, precision spectroscopy \cite{PRL_CQT_clock_Ba}, and quantum information processing \cite{APB_Akerman_2012,PRXQuantum_Tom_2022}.
The fundamental modification of the ideal linear Paul traps is the introduction of the endcap electrodes to achieve axial confinement. This alters the $a-q$ parameter space \cite{major2005charged} based on the choice of the driving scheme.
The resulting perturbations to the secular frequencies must therefore be carefully characterized, given the central role of the secular frequencies in determining the quantized motional dynamics of trapped ions \cite{Monroe_1995,Leibfried_2003}.

In another class of experiments, which investigate ion-atom interactions in hybrid traps, significantly larger LPT configurations are employed to ensure sufficient optical clearance for the laser beams used to co-trap the neutral atom cloud \cite{sduttamacro,ravi2012macro, APB_jyothi_2015,Felix_2012,Tomza_2018,Feldker_2020}.
These modifications introduce shifts to the secular frequencies of the trapped ions, which change the resonance conditions of the ions with specific e/m ratios.
Consequently, optimizing a LPT for targeted applications demands a comprehensive understanding of how trap geometry and specific electrical driving schemes influence these secular frequencies. While these effects have been analyzed in several foundational and recent works \cite{werthinstabilities,Filgueira_2025,oxfordthesis}, we experimentally investigate these deviations with dark Li$^+$ ions in a LPT geometry. We deliberately conduct the experiments at zero endcap voltage, which illustrates the effect of the presence of the endcap boundary. 

Sections \ref{sec:s2} and \ref{sec:s3} provide the required concepts and the experimental setup. Sections \ref{sec:s4} and \ref{sec:s5} deal with the investigations of trapped ions in axial and radial directions, respectively. Section \ref{sec:s6} provides a 2D map that utilizes the aspect ratios of three important trap dimensions to quantify the percentage shift of the LPTs from the ideal configuration. In section \ref{sec:s7} we propose a geometric modification to the LPT design to minimize the effects of endcaps while ensuring sufficient optical access.
\begin{figure}[h] 
    \centering 
    \includegraphics[width=0.47\textwidth]{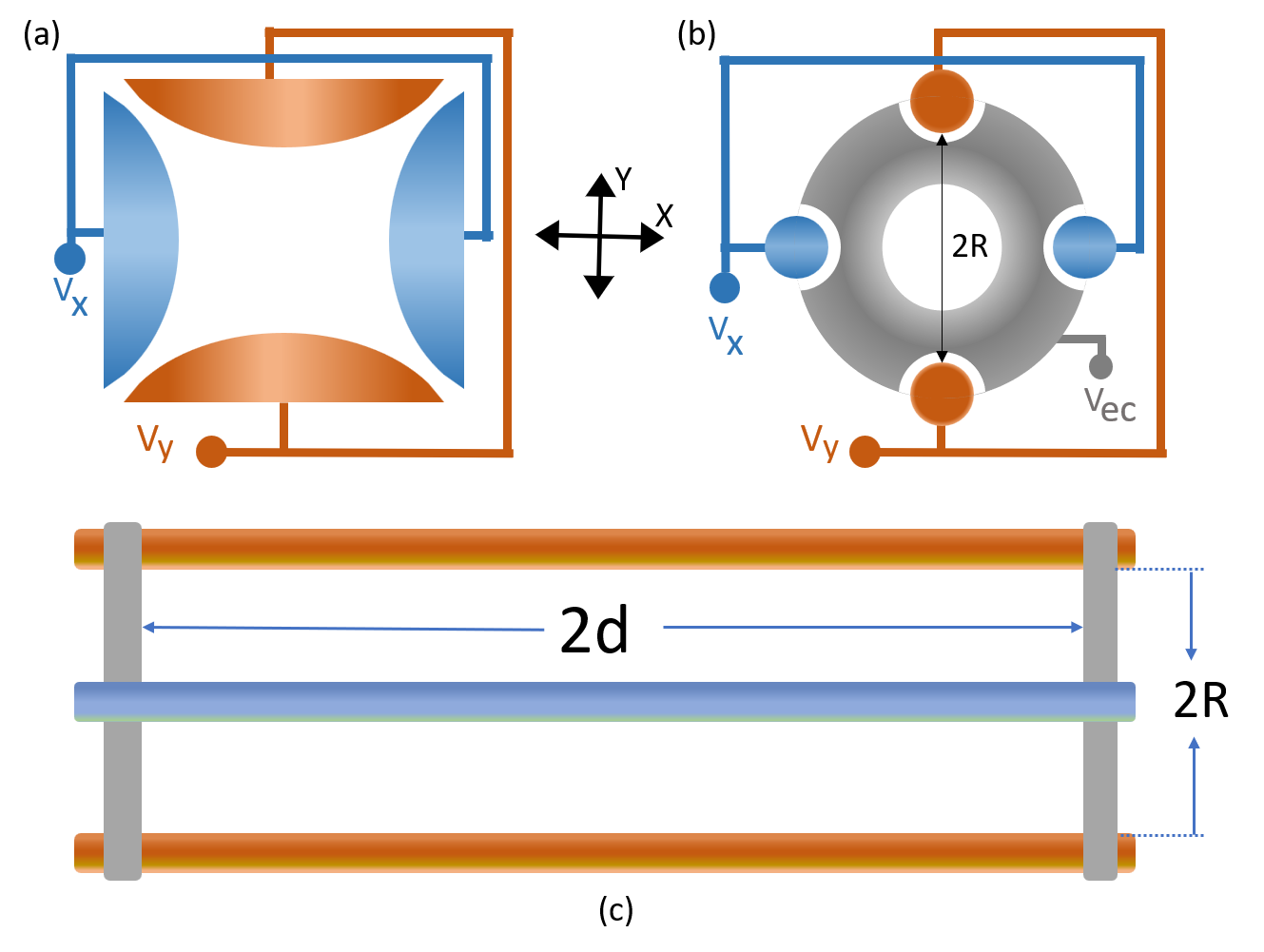} 
    \caption{(a) The ideal linear trap with hyperbolic boundaries. (b) A linear paul trap with linear cylindrical electrodes and hollow endcap electrodes. (c) Lateral view of the trap shown in (b). }
    \label{fig:NewTrapPics} 
\end{figure}
\vspace{-1.5cm}
\section{\label{sec:s2}Ideal LPT and Driving Schemes}
An ideal LPT schematic is shown in Fig. \ref{fig:NewTrapPics}(a). There are four electrodes (linear electrodes) having a hyperbolic boundary, placed equidistant from the Z-axis, the axis of the trap. R is the distance from the axis to the surface of these electrodes.
The trap can be operated in two different schemes, the Quadrupole Drive (QD) and Dipole Drive (DD) \cite{SCHWARTZ_2D_MS} by applying voltages V$_x$ and V$_y$ to the electrodes as shown in Fig. \ref{fig:NewTrapPics}(b). The biasing of the ion trap electrodes in QD and DD are given in the equation below. Here, $U_x$ and $U_y$ are the constant offset voltages applied to the electrodes. $V_{rf}$ and $\Omega_{rf}$ are the amplitude and the angular frequency of the RF drive, respectively.
\begin{align}
&\left.
\begin{aligned}
    V_x(t) &= U_x + V_{rf}\cos(\Omega_{rf} t) \\
    V_y(t) &= U_y - V_{rf}\cos(\Omega_{rf} t)
\end{aligned}
\right\}
\;\text{Quadrupole Drive}
\label{eq:QD}
\\[6pt]
&\left.
\begin{aligned}
    V_x(t) &= U_x + V_{rf}\cos(\Omega_{rf} t) \\
    V_y(t) &= U_y
\end{aligned}
\right\}
\;\text{Dipole Drive}
\label{eq:DD}
\end{align}
QD requires maintaining a phase difference of $\pi$ radians between the voltages applied to X and Y electrodes, Eq. (\ref{eq:QD}), while maintaining the same voltage amplitude on both. Hence, it is simpler to drive the LPT using the DD configuration \cite{Joger_2017,Meir_2018,Denschlag_2012,DD_Heinrich2018BeTrap,DD_Hempel2014Thesis,DD_Ballance2014}.

For an infinitely long set of linear electrodes of an ideal LPT, the analytical form of the radial potential in a QD configuration is given as
\cite{major2005charged}
\begin{multline}
   V(x,y,t) = V_{rf}Cos(\Omega_{rf} t) (\frac{x^2-y^2}{R^2} ) +\frac{(U_x + U_y)}{2}\\+ \frac{(U_x - U_y) (x^2-y^2)}{2 R^2}, \label{eq:VQD}
\end{multline}
whereas the analytical potential in DD configuration is 
\begin{multline}
   V(x,y,t) = \frac{V_{rf}}{2} Cos(\Omega_{rf} t)+
   \frac{V_{rf}}{2}Cos(\Omega_{rf} t) (\frac{x^2-y^2}{R^2} )\\ +\frac{(U_x + U_y)}{2}+ \frac{(U_x - U_y) (x^2-y^2)}{2 R^2}. \label{eq:VDD}
\end{multline}
We shall restrict the following discussion to cases where U$_x$ and U$_y$ are zero so that we steer clear of the effects of non-zero $a_{x,y}$ Mathieu parameters \cite{major2005charged} in the modification of secular frequencies.
In this work, we combine experiments with numerical simulation of trap operation using Simion \cite{simion} and Mathematica \cite{mathematica}.
The simulation studies on the dynamics of trapped Li$^+$ ions were performed for three trap configurations with different endcap separations listed in Table \ref{tab:traps3}, while the experiments were conducted using Trap 2 from the same table. For all configurations, the radial electrodes have a cylindrical cross-section with a radius of 3.5 mm and a separation between the radial electrodes of R = 12.25 mm.
\section{\label{sec:s3}Experimental platform}
The experimental platform is a hybrid ion-atom trap \cite{joshi_AI}.
It consists of a linear Paul trap (d = 27 mm, R = 12.4 mm, r = 1.75 mm) with two hollow endcap electrodes as shown in Fig. \ref{fig:NewTrapPics}(b,c).
The experiments here are done for the DD case, Eq. (\ref{eq:DD}). The V$_{rf}$ was set to 70 V, and the angular frequency $\Omega_{rf}$ was set to 2$\pi \times$1 MHz for all the experiments.
A dilute gas of Li cloud was formed using the technique of magneto-optical trapping (MOT). The Li$^+$ was formed by shining a blue LED onto the MOT that photoionizes the Li atoms from the $2P_{3/2}$ state \cite{joshi_AI}. 
These ions are held in the ion trap for specific time intervals and are detected by extracting them onto a micro-channel plate MCP for counting. Further details of the experimental apparatus will be reported separately. \cite{joshi_Instrumentation}.
\begin{table}[t]
\caption{Geometrical parameters of the three linear traps studied in this work. Trap 2 has the dimensions of the trap used in the experiment.}
\label{tab:traps3}
\begin{ruledtabular}
\begin{tabular}{lccc}
Parameter 
& Trap 1 
& Trap 2
& Trap 3 \\
\colrule
Endcap separation $2d$ (mm)    & 27  & 54  & 108 \\
\end{tabular}
\end{ruledtabular}
\end{table}
\section{\label{sec:s4}Effects on Axial Confinement}
The main difference between these two driving schemes for ideal LPT which is infinitely long and without endcap electrodes, is the potential along the trap axis ($x = y = 0$). The axial potential is constant at $\frac{(U_x + U_y)}{2}$ in the case of QD, Eq. (\ref{eq:VQD}). Whereas in the case of DD Eq. (\ref{eq:VDD}), the axial potential is $\frac{(U_x + U_y)}{2}+\frac{V_{rf}}{2} Cos(\Omega_{rf} t)$ which varies with time. This term does not have any spatial dependence for the ideal LPT and doesn't produce any force on ions.
\begin{figure}[b] 
    \centering 
    \includegraphics[width=0.48\textwidth]{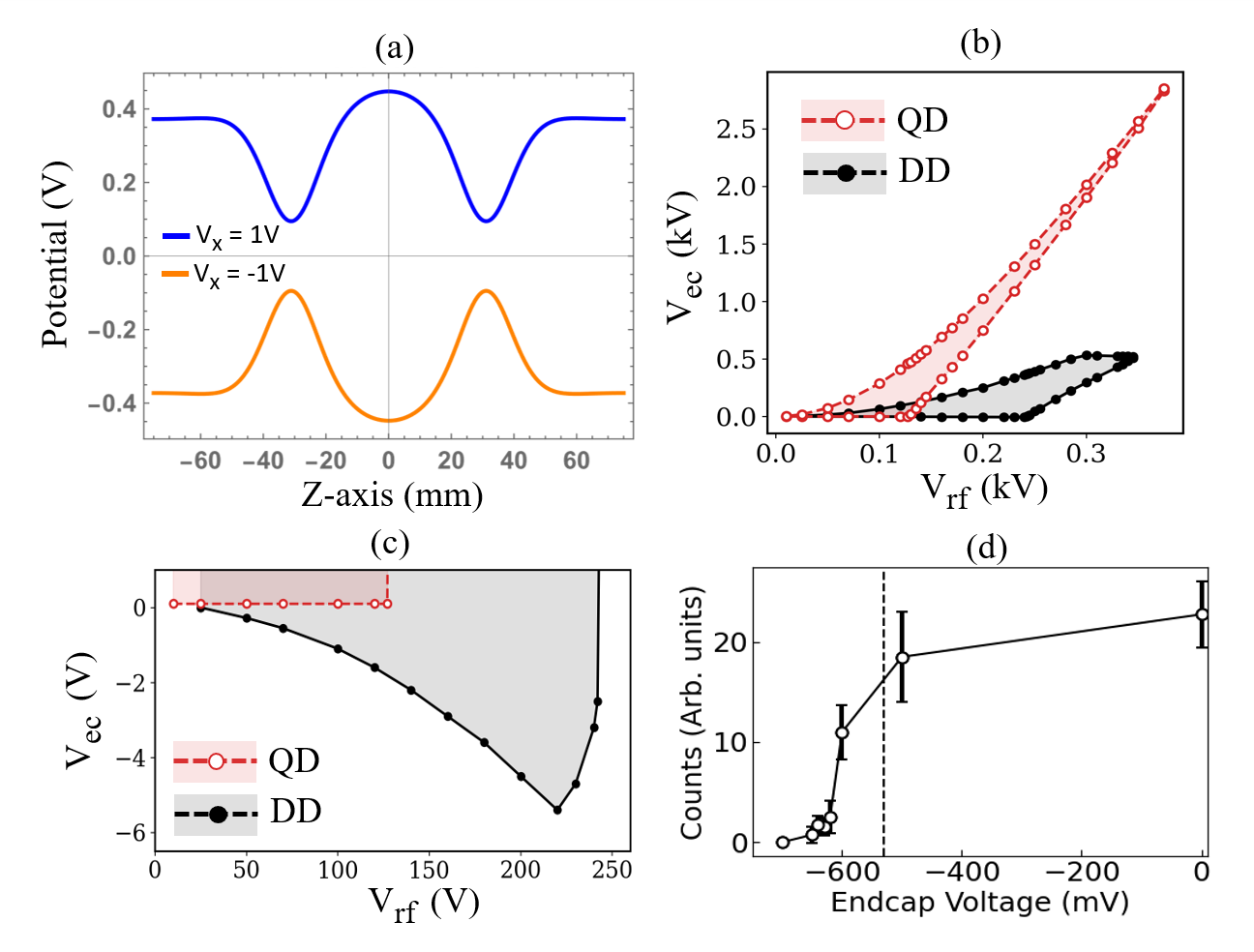} 
    \caption{Axial direction plots. a) Curvature of potential in DD case along the Z-axis. Here the Y-electrodes and the endcaps are grounded. b)The stability region in V$_{rf}$ vs V$_{ec}$ space for the DD and QD; the RF drive frequency was set to 2$\pi\times1$MHz and the simulation was done for Li$^+$ ions in Trap-2 of Table-1. c) The plot highlights the negative part of the stability region in (b). d) Experimental demonstration of trapping Li$^+$ ions with negative and zero endcap voltages. The dotted line is at the simulated voltage V$_{ec}$, Fig.\ref{fig:az_qZ}c, and represents the demarcation between stable and unstable simulated trajectories.} 
    \label{fig:az_qZ} 
\end{figure}
However, the LPTs employed in experiments are of finite length, and the confinement in the axial direction is achieved by introducing two coaxial electrodes normal to the trap axis, the end cap electrodes Fig. \ref{fig:NewTrapPics} (b,c). These electrodes are held at a constant positive (negative) voltage V$_{ec}$  to trap positive (negative) ions.
Consider the case where both the endcap electrodes are grounded and $\text{U}_x=\text{U}_y=0$.
In the case of QD, the potential along the axis ($x=y=0$) is constant and is zero. Hence, the axial trapping in the QD is achieved by applying a small voltage on the endcap electrodes. Therefore, the Mathieu parameter $a_z$ always has the same polarity as the charge of the trapped ions. 
However, in the case of DD, the potential at the centre of the trap oscillates with an amplitude $\simeq$V$_{rf}$/2 and frequency $\Omega_{rf}$, while the potential approaches zero at the endcap electrodes Fig. \ref{fig:az_qZ}(a).

This time-dependent potential along the Z-axis alters the entire stability region in $a_z$-$q_z$ space for DD.
Since the Mathieu parameters $a_z$ and $q_z$ are proportional to the constant V$_{ec}$ and time-dependent V$_{rf}$ voltage amplitudes\cite{major2005charged}, respectively, we present the simulation data of the stability region as a function of V$_{ec}$ and V$_{rf}$ in Fig. \ref{fig:az_qZ}(b). For the simulation , the ion trap (Trap-2 from Table-1) is operated with DD with a drive frequency of 2$\pi\times1$MHz. 
The ions start at the center of the trap with zero radial velocity and are iterated over ten different axial velocities with a spread of 100 m/s around zero.
The ion trajectories are computed for sufficient time (700 $\mu$s) such that their stability is unambiguously determined for different values of V$_{rf}$ and V$_{ec}$. The pairs of V$_{ec}$ and V$_{rf}$ for which the ions are trapped are shown in the shaded region (grey for DD and red for QD) in Fig. \ref{fig:az_qZ}(b).
\begin{figure}[b] 
    \centering 
    \includegraphics[width=0.45\textwidth]{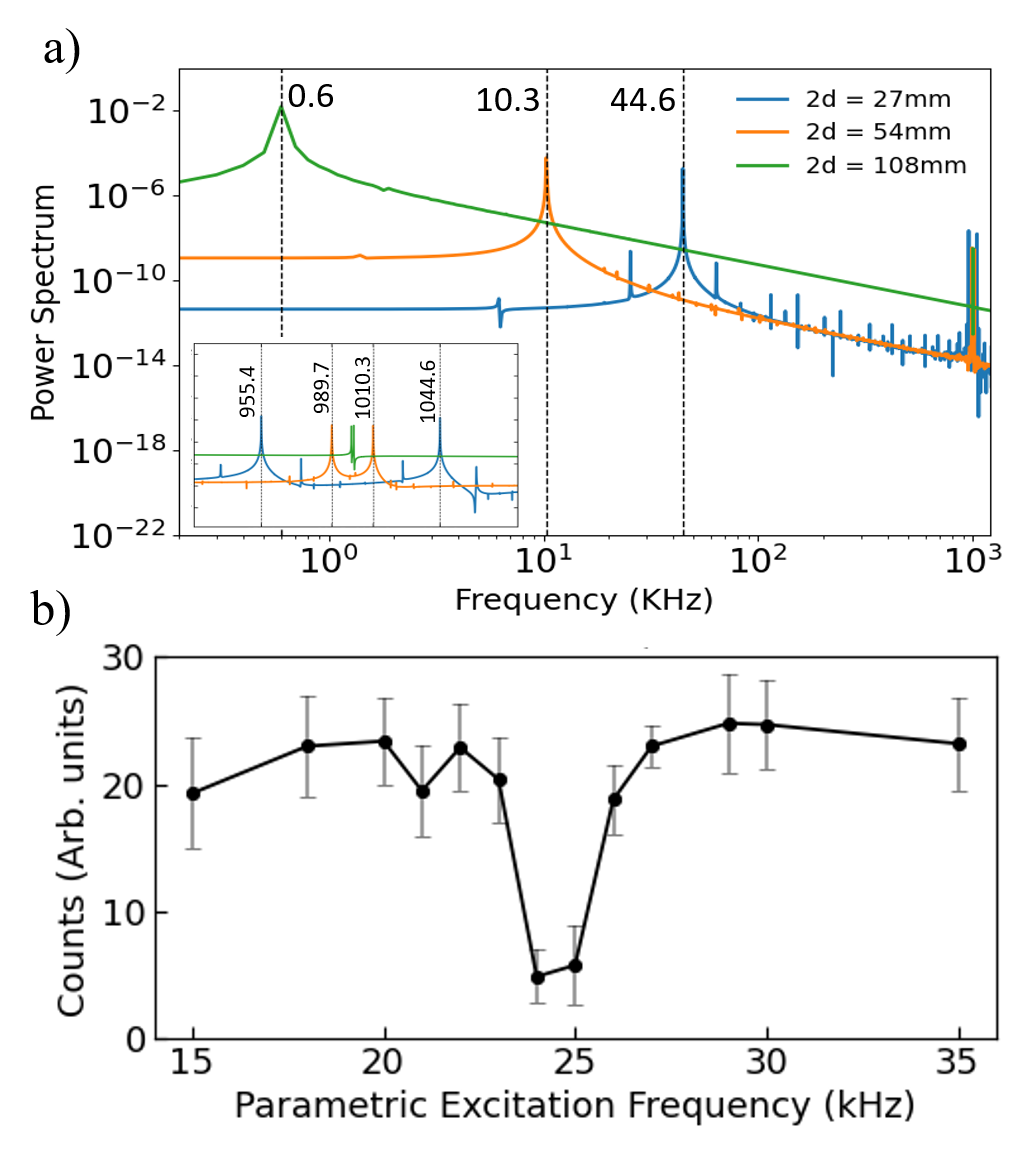}
    \caption{Simulations and experimental results for axial trapping of the Li$^+$ ions in DD configuration. (a) Power spectrum along the Z-axis for the Li$^+$ ions simulated for three different traps shown in Table \ref{tab:traps3}, where the endcaps are held at 0 Volts. The inset shows the micromotion peaks. (b) PE experiment along the axis of the trap. The vertical axis shows the number of ions that survived the PE drive. The horizontal axis shows the PE drive frequency. }
    \label{fig:pe_fz}
\end{figure}
For the QD case, the ions are unstable for V$_{ec} \leq 0$. On the other hand, for the DD case, because of the RF coupling to the axial field, a finite region of V$_{ec} \leq 0$ also results in stable trapping of ions as shown in Fig. \ref{fig:az_qZ}(c).

We successfully trapped the lithium ions with zero and negative endcap voltages. 
The trap was operated with V$_{rf}$ = 70 V and $\Omega_{rf} = 2\pi\times1$ MHz. The ions were loaded from the Li MOT that is created at the center of the ion trap. The ions were held in the trap for another 2.5 s before they were extracted onto the micro-channel plate for counting. The experiment was performed for various values of V$_{ec} \leq 0$. The experimental data in Fig. \ref{fig:az_qZ}(d) confirm the simulated signature, with the dotted line separating the stable and unstable regions.

We next present the determination of the axial secular frequency at zero endcap voltage. Fig. \ref{fig:pe_fz}(a) shows the simulated power spectrum of the trapped Li$^+$ ions for three traps of Table \ref{tab:traps3}. The secular frequency of Trap 2 of Table \ref{tab:traps3} is $\simeq$ 10.3 kHz. This was experimentally verified by the technique of parametric excitation (PE) \cite{PRA_Zhao_2002, APB_Schmidt_2020} by measuring the ion loss as seen in Fig. \ref{fig:pe_fz}(b).
The PE drive was given to both the endcap electrodes as V$_{ec}$ = 0 + V$^p_{ec}$Cos($\omega_p t$) Fig. \ref{fig:NewTrapPics}(b), where V$^p_{ec}$ was set to 200 mV to probe the axial resonance. The linear electrodes are operated in the DD configuration as specified in Eq. (\ref{eq:DD}) with Vrf = 70V and $\Omega_{rf}$ = $2\pi\times$1 MHz.
The ions were loaded for 0.5 s and were held in the presence of the parametric drive for another 2.5 s and then extracted onto the MCP. This experiment was performed for different values of the drive frequency $\omega_p $ ranging from 15 kHz to 35 kHz.
Fig. \ref{fig:pe_fz}(b) shows the number of ions that survive the parametric drive and are detected onto the MCP.
The dip in ion counts is at 24-25 kHz. The PE technique pumps maximum energy into the system at a frequency that is twice the resonant frequency of the system. Hence, the axial secular frequency is $\simeq 12$ kHz, and is in fair agreement with the simulation. The slight disagreement between the simulation and the experimental result is attributed to a combination of two reasons. First is the difference in the spacing of the simulated electrodes (R = 12.25 mm) and the experimental assembly (R = 12.4 mm) due to grid constraints in Simion \cite{simion}, and second is the actual RF voltage amplitude on the electrodes compared to the set value during the experiment.

\begin{figure}[h]
    \centering
    \includegraphics[width=0.48\textwidth]{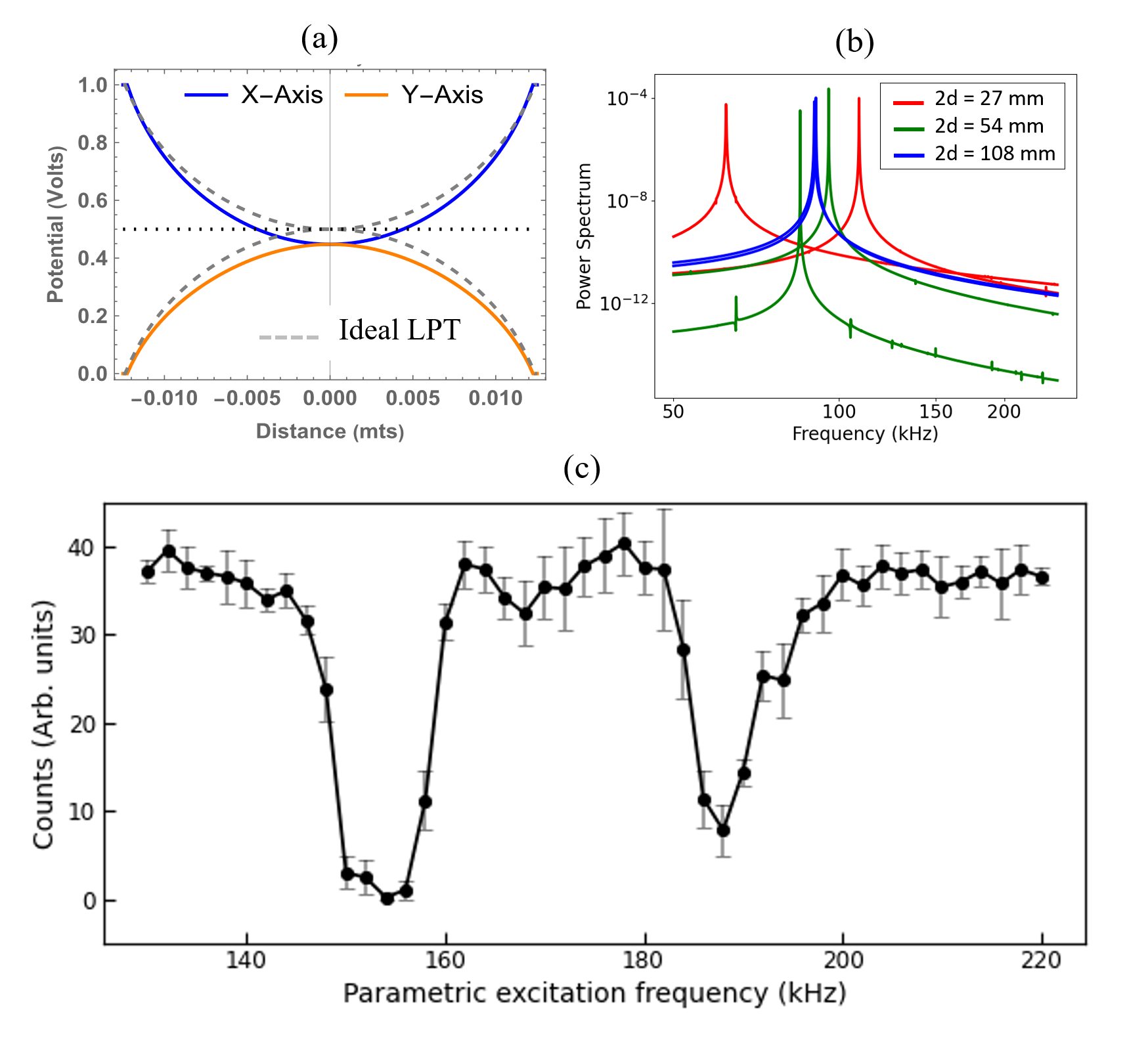}
    \caption{Radial Axis plots (a) Curvature of the potential in DD case along X(blue) and Y(orange) Axes for V$_x$ = 1V and V$_y$ = 0V. The grey dashed lines are the curvatures for an ideal LPT, and the dotted horizontal line is at 0.5 V. (b)The power spectrum in radial axes simulated for three different traps of Table \ref{tab:traps3} with different endcap separations. (c) The PE experiment for the Li$^+$ ions along the radial axes for endcap voltage of 0 volts.}
    \label{fig:radialplots}
\end{figure}
\section{\label{sec:s5}Effects on radial confinement}
The physical presence of the endcaps will also affect the curvatures along the radial directions.
Consider the DD case with grounded endcap electrodes. The Y-axis electrodes are grounded, and the X-axis electrodes oscillate with the voltage V$_{rf}$ and frequency $\Omega_{rf}$. Fig. \ref{fig:radialplots}(a) shows the plot for when the voltage on the X-axis electrodes is 1V. The potential at the center, in the presence of the endcap electrodes, is less than V$_{rf}$(t)/2. 
This leads to unequal curvatures along the X and Y axes as seen in Fig.\ref{fig:radialplots}(a).
This results in unequal $q$ parameters in the radial direction ($q_x \neq $ $q_y$), leading to different radial secular frequencies.
The Fourier transform plots for the trapped ion trajectories are shown in Fig.\ref{fig:radialplots}(b) for the three traps of Table \ref{tab:traps3}.
We validate our results from simulations with an experimental result using the same method of PE described earlier to determine the secular frequencies of the trapped ions.
The parametric drive was given as specified in the equation below.
\begin{align}
V_x(t) &= (70 \ \text{V}) \cos(2\pi \times (\text{1MHz}) t) + V_p \cos(\omega_p t), \\
V_y(t) &= 0 \text{V}, \\
V_{ec}(t) &= 0  \text{V} .
\end{align}
Here V$_p$ was set to 140 mV. The ion trap was switched on, followed by loading of the ions into the trap for 0.1 sec. The ions were held in the trap for another 1.5 sec and were extracted onto the MCP.
This was done for different values of $\omega_p$ as can be seen in Fig. \ref{fig:radialplots}(c). The two resonances in the figure are at 2$\omega_y$ = 158 kHz and 2$\omega_x$ = 184 kHz.

\begin{figure}[h] 
    \centering 
    \includegraphics[width=0.45\textwidth]{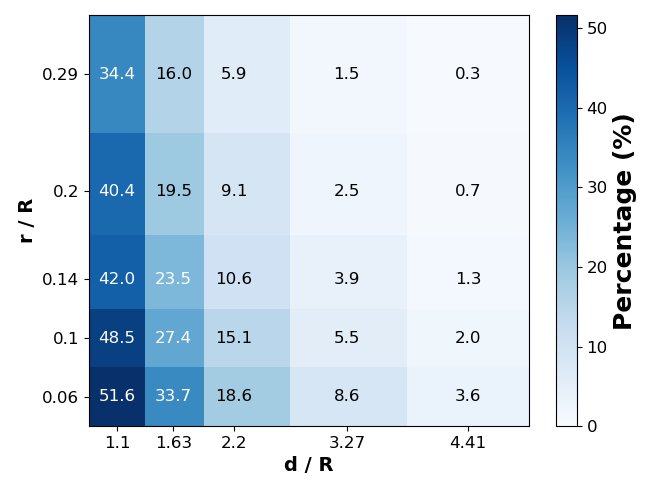} 
    \caption{Percentage deviation of the potential at the center for dipole drive from its ideal value listed for different $r/R$ and $d/R$ ratios. One of the RF electrode pairs was set to 1 V, and the rest of the electrodes were grounded.} 
    \label{fig:aspect ratio} 
\end{figure}
\section{\label{sec:s6}Categorising the LPT\lowercase{s}}
All the effects of the DD can be mapped to one phenomenon, the potential at the centre of the trap. For the ideal LPT in DD, the potential at the center is V$_{rf}/2$. Deviations from this value serve as a measure of all the effects discussed above.
Hence, we present a scaling map that gives the percentage shift of the potential at the trap centre from its ideal value.
Any linear four-pole trap has three important length parameters: the separation between the endcaps (2d), the separation between the linear electrodes (2R) and the size of the linear electrodes (2r) as defined in Fig. \ref{fig:NewTrapPics}. 
\begin{figure}[h] 
    \centering 
    \includegraphics[width=0.45\textwidth]{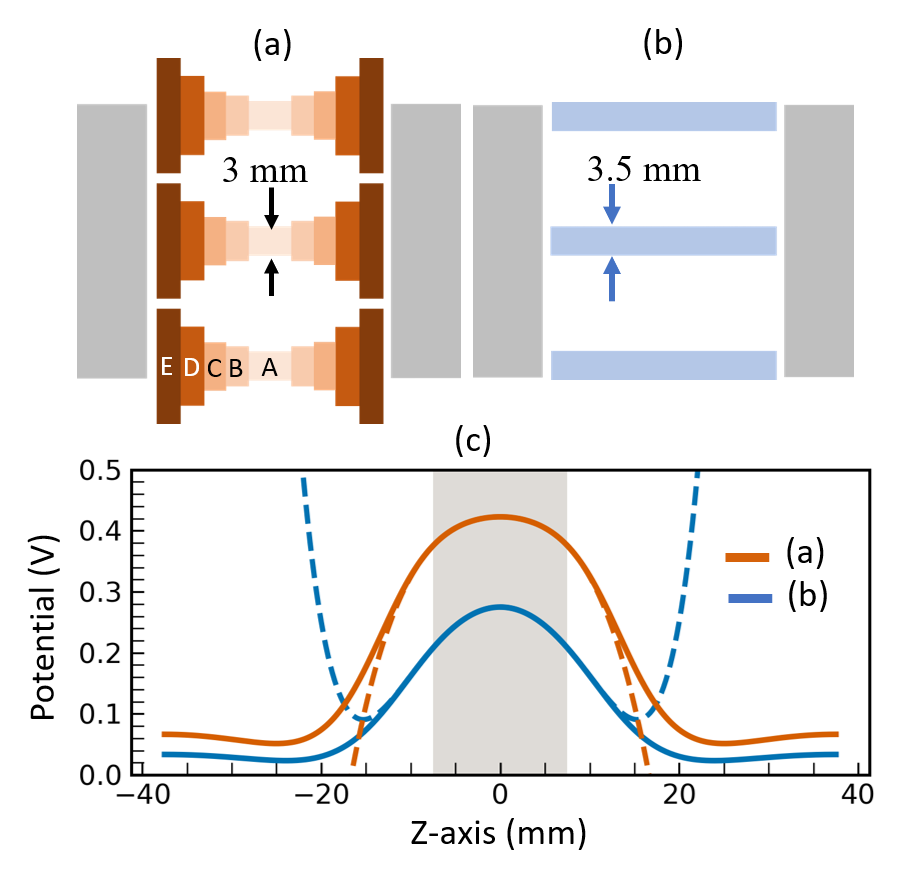} 
    \caption{(a) Lateral view of the new design with tapered linear electrodes. Each linear electrode has a tapered cross-section with segments A-E having different radii as follows: A = 1.5 mm, B = 2 mm, C = 2.5 mm, D = 4.25 mm and E = 6.25 mm. The lengths of the segments A and B-E are 5 mm and 2.5 mm, respectively. (b) A simple LPT with r = 1.75 mm and with the same endcap separation $2d$ and R as the new design in (a). (c) The axial potential in the case of 0 V endcap. The orange and the blue curves are for trap designs (a) and (b) respectively. This clearly shows a comparatively flatter curvature at the center. The grey shaded region is used to fit a polynomial to compare the flatness of the potentials.} 
    \label{fig:newtrap} 
\end{figure}
The percentage shift for different aspect ratios is given in Fig. \ref{fig:aspect ratio}.
This quantification was done by categorising the LPTs in terms of two aspect ratios d/R and r/R.
This percentage shift is in theory valid for any length scale and is independent of any particular values of R, r, and d. It is computed at the trap centre for the configuration where V$_{ec}$=0, one pair of linear electrodes is grounded, and the other is pair is set to 1V. The simulations for refining the potentials are done in Simion \cite{simion}.
This type of categorizing is beneficial while scaling the ion trap architectures, where the cost of modified electrode geometries needs to be minimized while miniaturizing the traps and maximizing optical access. We shall use this map to propose a minor modification to LPT that greatly reduces the radial-to-axial coupling of the potential.
Note that categorising the ion traps using these two aspect ratios was done for cylindrical cross-sections of the linear electrodes and hollow endcap electrodes as shown in Fig. \ref{fig:NewTrapPics}(b). The percentage values will slightly differ for other geometries.

\section{\label{sec:s7}Design to Flatten Axial Potential}
The inset of Fig. \ref{fig:pe_fz}(a) shows peaks at $\Omega_{rf}\pm\omega_z$. These are the intrinsic axial micromotion peaks due to the time-dependent potential along the trap axis. Therefore, the DD allows only a point node at the center of the trap, in contrast with the QD, which retains the advantage of a nodal line along the trap axis. The effect of the axial micromotion in DD can be minimized by flattening the axial potential at zero endcap voltage. This would require increasing both $d/R$ and $r/R$ ratios, as can be seen from Fig \ref{fig:aspect ratio} which hinders optical access and increases the overall size of the trap. Below, we propose a modification to flatten the potential considerably while retaining the optical access.

Note that the categorization in Fig.\ref{fig:aspect ratio} is done for the uniform cross-sections of the electrodes across the length of the trap. As a result, the curvature of the axial potential gets adjusted along with the potential at the trap center. However, adjusting the electrode sizes away from the trap center tunes the flatness of the axial potential while retaining the optical access.
The design illustrated in Fig.~\ref{fig:newtrap}(a), shields the endcaps while preserving both optical access at the trap center and the characteristic trap size (the $d/R$ ratio). 

At the trap center, the ratios ($r/R$) and ($d/R$) are $0.14$ and $1.1$, respectively. However, to achieve shielding, the radius of all four linear electrodes is progressively increased (tapered) with an increase in axial distance from the trap center. For comparison, Fig.~\ref{fig:newtrap}(b) depicts a standard linear Paul trap (LPT) of identical aspect ratios but with uniform electrode radii throughout, which exhibits a high radial-to-axial coupling of 42\% (see Fig. \ref{fig:aspect ratio}). 

The simulated axial potentials for both designs are plotted in Fig.~\ref{fig:newtrap}(c), where the voltage on one pair of linear electrodes is set to 1~V, while the endcaps and the remaining electrode pair are grounded. The potential at the center of the proposed tapered trap is 0.42~V, a 16\% deviation from the ideal value of 0.5~V. This represents a significant reduction in radial-to-axial coupling compared to the 42\% deviation observed in the uniform geometry. 

The axial potential profile in Fig.\ref{fig:newtrap}(c) of the new design is noticeably flatter near the center as expected.
The region shaded in grey in Fig.\ref{fig:newtrap}(c) was used to fit an even-order polynomial series $\Sigma a_nx^n$ up to the 6$^{\text{th}}$ order. The second-order coefficient $a_2$ for the new trap design is reduced by a factor of 2 compared to the conventional design. 
Hence, the curvature of the potential at the centre is relatively flattened compared to the regular LPT design.  
This flattened potential reduces the axial force on the trapped ions and hence the $q_z$ parameter.  The simulated secular frequencies ($\omega_x$, $\omega_y$, $\omega_z$) in DD (V$_{rf}$ = 70 V, $\Omega_{rf} = 2\pi \times $1 MHz) at 0 V endcap voltage for the proposed trap design (Fig.\ref{fig:newtrap}a) and the conventional design (Fig.\ref{fig:newtrap}b) are (135, 109, 23) KHz and (110, 64, 45) KHz respectively. There is a significant reduction in the axial secular frequency and also in the split between radial secular frequencies. 

Many LPTs and their modifications, such as blade traps, use the DD to drive the ion trap for various experimental studies \cite{Jyothi_2019,Schmidt_2020,DD_Used_CJBalance_PRL,DD_used_Joshi2020PGC,DD_used_Kaewuam_2018,DD_used_Olmschenk}.
While there are active efforts to build robust QD electronics \cite{Surendra_2026}, the proposed tapered architecture offers an added passive solution at the design stage to suppress radial-to-axial RF coupling without compromising optical access or altering the size of the trap.
\section{\label{sec:s8}conclusions}

We have investigated the combined influence of the DD scheme and the aspect ratios of the trap electrodes on the properties of trapped ions in LPTs. We experimentally demonstrate the trapping of Li$^+$ ions at both zero and negative endcap voltages, showing that axial confinement can be sustained even in the absence of conventional positive endcap voltages. 
We further show that the DD configuration breaks the radial symmetry of the trapping potential, leading to unequal radial secular frequencies.
DD is the source of axial micro-motion that cannot be eliminated, but it can be substantially reduced through appropriate trap design.
Finally, we propose a modified trap architecture that preserves both the trap dimensions and the optical access while significantly suppressing radial-to-axial RF coupling and flattening the axial potential. These results provide practical design guidelines for optimizing dipole-driven linear Paul traps in applications requiring enhanced optical accessibility while maintaining the trap sizes.
\nocite{*}


\begin{thebibliography}{40}%
\makeatletter
\providecommand \@ifxundefined [1]{%
 \@ifx{#1\undefined}
}%
\providecommand \@ifnum [1]{%
 \ifnum #1\expandafter \@firstoftwo
 \else \expandafter \@secondoftwo
 \fi
}%
\providecommand \@ifx [1]{%
 \ifx #1\expandafter \@firstoftwo
 \else \expandafter \@secondoftwo
 \fi
}%
\providecommand \natexlab [1]{#1}%
\providecommand \enquote  [1]{``#1''}%
\providecommand \bibnamefont  [1]{#1}%
\providecommand \bibfnamefont [1]{#1}%
\providecommand \citenamefont [1]{#1}%
\providecommand \href@noop [0]{\@secondoftwo}%
\providecommand \href [0]{\begingroup \@sanitize@url \@href}%
\providecommand \@href[1]{\@@startlink{#1}\@@href}%
\providecommand \@@href[1]{\endgroup#1\@@endlink}%
\providecommand \@sanitize@url [0]{\catcode `\\12\catcode `\$12\catcode `\&12\catcode `\#12\catcode `\^12\catcode `\_12\catcode `\%12\relax}%
\providecommand \@@startlink[1]{}%
\providecommand \@@endlink[0]{}%
\providecommand \url  [0]{\begingroup\@sanitize@url \@url }%
\providecommand \@url [1]{\endgroup\@href {#1}{\urlprefix }}%
\providecommand \urlprefix  [0]{URL }%
\providecommand \Eprint [0]{\href }%
\providecommand \doibase [0]{https://doi.org/}%
\providecommand \selectlanguage [0]{\@gobble}%
\providecommand \bibinfo  [0]{\@secondoftwo}%
\providecommand \bibfield  [0]{\@secondoftwo}%
\providecommand \translation [1]{[#1]}%
\providecommand \BibitemOpen [0]{}%
\providecommand \bibitemStop [0]{}%
\providecommand \bibitemNoStop [0]{.\EOS\space}%
\providecommand \EOS [0]{\spacefactor3000\relax}%
\providecommand \BibitemShut  [1]{\csname bibitem#1\endcsname}%
\let\auto@bib@innerbib\@empty
\bibitem [{\citenamefont {Douglas}\ \emph {et~al.}(2004)\citenamefont {Douglas}, \citenamefont {Frank},\ and\ \citenamefont {Mao}}]{Douglas}%
  \BibitemOpen
  \bibfield  {author} {\bibinfo {author} {\bibfnamefont {D.~J.}\ \bibnamefont {Douglas}}, \bibinfo {author} {\bibfnamefont {A.~J.}\ \bibnamefont {Frank}},\ and\ \bibinfo {author} {\bibfnamefont {D.}~\bibnamefont {Mao}},\ }\bibfield  {title} {\bibinfo {title} {Linear ion traps in mass spectrometry},\ }\href {https://doi.org/https://doi.org/10.1002/mas.20004} {\bibfield  {journal} {\bibinfo  {journal} {Mass Spectrometry Reviews}\ }\textbf {\bibinfo {volume} {24}},\ \bibinfo {pages} {1} (\bibinfo {year} {2004})}\BibitemShut {NoStop}%
\bibitem [{\citenamefont {Kitaoka}\ \emph {et~al.}(2013)\citenamefont {Kitaoka}, \citenamefont {Yoshida}, \citenamefont {Yamamoto}, \citenamefont {Jung},\ and\ \citenamefont {Hasegawa}}]{Kitaoka_2013}%
  \BibitemOpen
  \bibfield  {author} {\bibinfo {author} {\bibfnamefont {M.}~\bibnamefont {Kitaoka}}, \bibinfo {author} {\bibfnamefont {T.}~\bibnamefont {Yoshida}}, \bibinfo {author} {\bibfnamefont {Y.}~\bibnamefont {Yamamoto}}, \bibinfo {author} {\bibfnamefont {K.}~\bibnamefont {Jung}},\ and\ \bibinfo {author} {\bibfnamefont {S.}~\bibnamefont {Hasegawa}},\ }\bibfield  {title} {\bibinfo {title} {Trapping and laser cooling of trace ca+ isotopes injected from an inductively coupled plasma mass spectrometer},\ }\href {https://doi.org/10.1039/C3JA00004D} {\bibfield  {journal} {\bibinfo  {journal} {J. Anal. At. Spectrom.}\ }\textbf {\bibinfo {volume} {28}},\ \bibinfo {pages} {1292} (\bibinfo {year} {2013})}\BibitemShut {NoStop}%
\bibitem [{\citenamefont {Sullivan}\ \emph {et~al.}(2011)\citenamefont {Sullivan}, \citenamefont {Rellergert}, \citenamefont {Kotochigova}, \citenamefont {Chen}, \citenamefont {Schowalter},\ and\ \citenamefont {Hudson}}]{Hudson}%
  \BibitemOpen
  \bibfield  {author} {\bibinfo {author} {\bibfnamefont {S.~T.}\ \bibnamefont {Sullivan}}, \bibinfo {author} {\bibfnamefont {W.~G.}\ \bibnamefont {Rellergert}}, \bibinfo {author} {\bibfnamefont {S.}~\bibnamefont {Kotochigova}}, \bibinfo {author} {\bibfnamefont {K.}~\bibnamefont {Chen}}, \bibinfo {author} {\bibfnamefont {S.~J.}\ \bibnamefont {Schowalter}},\ and\ \bibinfo {author} {\bibfnamefont {E.~R.}\ \bibnamefont {Hudson}},\ }\bibfield  {title} {\bibinfo {title} {Trapping molecular ions formed via photo-associative ionization of ultracold atoms},\ }\href {https://doi.org/10.1039/C1CP21205B} {\bibfield  {journal} {\bibinfo  {journal} {Phys. Chem. Chem. Phys.}\ }\textbf {\bibinfo {volume} {13}},\ \bibinfo {pages} {18859} (\bibinfo {year} {2011})}\BibitemShut {NoStop}%
\bibitem [{\citenamefont {H{\"a}rter}\ and\ \citenamefont {Hecker~Denschlag}(2014)}]{denschlag2014macro}%
  \BibitemOpen
  \bibfield  {author} {\bibinfo {author} {\bibfnamefont {A.}~\bibnamefont {H{\"a}rter}}\ and\ \bibinfo {author} {\bibfnamefont {J.}~\bibnamefont {Hecker~Denschlag}},\ }\bibfield  {title} {\bibinfo {title} {Cold atom--ion experiments in hybrid traps},\ }\href@noop {} {\bibfield  {journal} {\bibinfo  {journal} {Contemporary Physics}\ }\textbf {\bibinfo {volume} {55}},\ \bibinfo {pages} {33} (\bibinfo {year} {2014})}\BibitemShut {NoStop}%
\bibitem [{\citenamefont {Smith}\ \emph {et~al.}(2005)\citenamefont {Smith}, \citenamefont {Makarov},\ and\ \citenamefont {Lin}}]{jian2005macro}%
  \BibitemOpen
  \bibfield  {author} {\bibinfo {author} {\bibfnamefont {W.~W.}\ \bibnamefont {Smith}}, \bibinfo {author} {\bibfnamefont {O.~P.}\ \bibnamefont {Makarov}},\ and\ \bibinfo {author} {\bibfnamefont {J.}~\bibnamefont {Lin}},\ }\bibfield  {title} {\bibinfo {title} {Cold ion--neutral collisions in a hybrid trap},\ }\href@noop {} {\bibfield  {journal} {\bibinfo  {journal} {Journal of Modern Optics}\ }\textbf {\bibinfo {volume} {52}},\ \bibinfo {pages} {2253} (\bibinfo {year} {2005})}\BibitemShut {NoStop}%
\bibitem [{\citenamefont {Arnold}\ \emph {et~al.}(2020)\citenamefont {Arnold}, \citenamefont {Kaewuam}, \citenamefont {Chanu}, \citenamefont {Tan}, \citenamefont {Zhang},\ and\ \citenamefont {Barrett}}]{PRL_CQT_clock_Ba}%
  \BibitemOpen
  \bibfield  {author} {\bibinfo {author} {\bibfnamefont {K.~J.}\ \bibnamefont {Arnold}}, \bibinfo {author} {\bibfnamefont {R.}~\bibnamefont {Kaewuam}}, \bibinfo {author} {\bibfnamefont {S.~R.}\ \bibnamefont {Chanu}}, \bibinfo {author} {\bibfnamefont {T.~R.}\ \bibnamefont {Tan}}, \bibinfo {author} {\bibfnamefont {Z.}~\bibnamefont {Zhang}},\ and\ \bibinfo {author} {\bibfnamefont {M.~D.}\ \bibnamefont {Barrett}},\ }\bibfield  {title} {\bibinfo {title} {Precision measurements of the $^{138}{\mathrm{ba}}^{+}$ $6s{^{2}S}_{1/2}\ensuremath{-}5d{^{2}D}_{5/2}$ clock transition},\ }\href {https://doi.org/10.1103/PhysRevLett.124.193001} {\bibfield  {journal} {\bibinfo  {journal} {Phys. Rev. Lett.}\ }\textbf {\bibinfo {volume} {124}},\ \bibinfo {pages} {193001} (\bibinfo {year} {2020})}\BibitemShut {NoStop}%
\bibitem [{\citenamefont {Akerman}\ \emph {et~al.}(2012)\citenamefont {Akerman}, \citenamefont {Glickman},\ and\ \citenamefont {Kotler}}]{APB_Akerman_2012}%
  \BibitemOpen
  \bibfield  {author} {\bibinfo {author} {\bibfnamefont {N.}~\bibnamefont {Akerman}}, \bibinfo {author} {\bibfnamefont {Y.}~\bibnamefont {Glickman}},\ and\ \bibinfo {author} {\bibfnamefont {S.}~\bibnamefont {Kotler}},\ }\bibfield  {title} {\bibinfo {title} {Quantum control of 88sr+ in a miniature linear paul trap},\ }\href {https://doi.org/10.1007/s00340-011-4807-6} {\bibfield  {journal} {\bibinfo  {journal} {Applied physics. B}\ }\textbf {\bibinfo {volume} {107}} (\bibinfo {year} {2012})}\BibitemShut {NoStop}%
\bibitem [{\citenamefont {Manovitz}\ \emph {et~al.}(2022)\citenamefont {Manovitz}, \citenamefont {Shapira}, \citenamefont {Gazit}, \citenamefont {Akerman},\ and\ \citenamefont {Ozeri}}]{PRXQuantum_Tom_2022}%
  \BibitemOpen
  \bibfield  {author} {\bibinfo {author} {\bibfnamefont {T.}~\bibnamefont {Manovitz}}, \bibinfo {author} {\bibfnamefont {Y.}~\bibnamefont {Shapira}}, \bibinfo {author} {\bibfnamefont {L.}~\bibnamefont {Gazit}}, \bibinfo {author} {\bibfnamefont {N.}~\bibnamefont {Akerman}},\ and\ \bibinfo {author} {\bibfnamefont {R.}~\bibnamefont {Ozeri}},\ }\bibfield  {title} {\bibinfo {title} {Trapped-ion quantum computer with robust entangling gates and quantum coherent feedback},\ }\href {https://doi.org/10.1103/PRXQuantum.3.010347} {\bibfield  {journal} {\bibinfo  {journal} {PRX Quantum}\ }\textbf {\bibinfo {volume} {3}},\ \bibinfo {pages} {010347} (\bibinfo {year} {2022})}\BibitemShut {NoStop}%
\bibitem [{\citenamefont {Major}\ \emph {et~al.}(2005)\citenamefont {Major}, \citenamefont {Gheorghe},\ and\ \citenamefont {Werth}}]{major2005charged}%
  \BibitemOpen
  \bibfield  {author} {\bibinfo {author} {\bibfnamefont {F.~G.}\ \bibnamefont {Major}}, \bibinfo {author} {\bibfnamefont {V.~N.}\ \bibnamefont {Gheorghe}},\ and\ \bibinfo {author} {\bibfnamefont {G.}~\bibnamefont {Werth}},\ }\href@noop {} {\emph {\bibinfo {title} {Charged particle traps: physics and techniques of charged particle field confinement}}},\ Vol.~\bibinfo {volume} {37}\ (\bibinfo  {publisher} {Springer Science \& Business Media},\ \bibinfo {year} {2005})\BibitemShut {NoStop}%
\bibitem [{\citenamefont {Monroe}\ \emph {et~al.}(1995)\citenamefont {Monroe}, \citenamefont {Meekhof}, \citenamefont {King}, \citenamefont {Jefferts}, \citenamefont {Itano}, \citenamefont {Wineland},\ and\ \citenamefont {Gould}}]{Monroe_1995}%
  \BibitemOpen
  \bibfield  {author} {\bibinfo {author} {\bibfnamefont {C.}~\bibnamefont {Monroe}}, \bibinfo {author} {\bibfnamefont {D.~M.}\ \bibnamefont {Meekhof}}, \bibinfo {author} {\bibfnamefont {B.~E.}\ \bibnamefont {King}}, \bibinfo {author} {\bibfnamefont {S.~R.}\ \bibnamefont {Jefferts}}, \bibinfo {author} {\bibfnamefont {W.~M.}\ \bibnamefont {Itano}}, \bibinfo {author} {\bibfnamefont {D.~J.}\ \bibnamefont {Wineland}},\ and\ \bibinfo {author} {\bibfnamefont {P.}~\bibnamefont {Gould}},\ }\bibfield  {title} {\bibinfo {title} {Resolved-sideband raman cooling of a bound atom to the 3d zero-point energy},\ }\href {https://doi.org/10.1103/PhysRevLett.75.4011} {\bibfield  {journal} {\bibinfo  {journal} {Phys. Rev. Lett.}\ }\textbf {\bibinfo {volume} {75}},\ \bibinfo {pages} {4011} (\bibinfo {year} {1995})}\BibitemShut {NoStop}%
\bibitem [{\citenamefont {Leibfried}\ \emph {et~al.}(2003)\citenamefont {Leibfried}, \citenamefont {Blatt}, \citenamefont {Monroe},\ and\ \citenamefont {Wineland}}]{Leibfried_2003}%
  \BibitemOpen
  \bibfield  {author} {\bibinfo {author} {\bibfnamefont {D.}~\bibnamefont {Leibfried}}, \bibinfo {author} {\bibfnamefont {R.}~\bibnamefont {Blatt}}, \bibinfo {author} {\bibfnamefont {C.}~\bibnamefont {Monroe}},\ and\ \bibinfo {author} {\bibfnamefont {D.}~\bibnamefont {Wineland}},\ }\bibfield  {title} {\bibinfo {title} {Quantum dynamics of single trapped ions},\ }\href {https://doi.org/10.1103/RevModPhys.75.281} {\bibfield  {journal} {\bibinfo  {journal} {Rev. Mod. Phys.}\ }\textbf {\bibinfo {volume} {75}},\ \bibinfo {pages} {281} (\bibinfo {year} {2003})}\BibitemShut {NoStop}%
\bibitem [{\citenamefont {Rahaman}\ \emph {et~al.}(2024)\citenamefont {Rahaman}, \citenamefont {Baidya},\ and\ \citenamefont {Dutta}}]{sduttamacro}%
  \BibitemOpen
  \bibfield  {author} {\bibinfo {author} {\bibfnamefont {B.}~\bibnamefont {Rahaman}}, \bibinfo {author} {\bibfnamefont {S.}~\bibnamefont {Baidya}},\ and\ \bibinfo {author} {\bibfnamefont {S.}~\bibnamefont {Dutta}},\ }\bibfield  {title} {\bibinfo {title} {{A versatile apparatus for simultaneous trapping of multiple species of ultracold atoms and ions to enable studies of low energy collisions and cold chemistry}},\ }\href {https://doi.org/10.1063/5.0193481} {\bibfield  {journal} {\bibinfo  {journal} {The Journal of Chemical Physics}\ }\textbf {\bibinfo {volume} {160}},\ \bibinfo {pages} {064201} (\bibinfo {year} {2024})}\BibitemShut {NoStop}%
\bibitem [{\citenamefont {Ravi}\ \emph {et~al.}(2012)\citenamefont {Ravi}, \citenamefont {Lee}, \citenamefont {Sharma}, \citenamefont {Werth},\ and\ \citenamefont {Rangwala}}]{ravi2012macro}%
  \BibitemOpen
  \bibfield  {author} {\bibinfo {author} {\bibfnamefont {K.}~\bibnamefont {Ravi}}, \bibinfo {author} {\bibfnamefont {S.}~\bibnamefont {Lee}}, \bibinfo {author} {\bibfnamefont {A.}~\bibnamefont {Sharma}}, \bibinfo {author} {\bibfnamefont {G.}~\bibnamefont {Werth}},\ and\ \bibinfo {author} {\bibfnamefont {S.}~\bibnamefont {Rangwala}},\ }\bibfield  {title} {\bibinfo {title} {Combined ion and atom trap for low-temperature ion--atom physics},\ }\href@noop {} {\bibfield  {journal} {\bibinfo  {journal} {Applied Physics B}\ }\textbf {\bibinfo {volume} {107}},\ \bibinfo {pages} {971} (\bibinfo {year} {2012})}\BibitemShut {NoStop}%
\bibitem [{\citenamefont {Jyothi}\ \emph {et~al.}(2015)\citenamefont {Jyothi}, \citenamefont {Ray},\ and\ \citenamefont {Rangwala}}]{APB_jyothi_2015}%
  \BibitemOpen
  \bibfield  {author} {\bibinfo {author} {\bibfnamefont {S.}~\bibnamefont {Jyothi}}, \bibinfo {author} {\bibfnamefont {T.}~\bibnamefont {Ray}},\ and\ \bibinfo {author} {\bibfnamefont {S.}~\bibnamefont {Rangwala}},\ }\bibfield  {title} {\bibinfo {title} {Phase-sensitive radial extraction and mass spectrometry of trapped ions in a compact geometry},\ }\href {https://doi.org/10.1007/s00340-014-5961-4} {\bibfield  {journal} {\bibinfo  {journal} {Applied Bhysics B}\ }\textbf {\bibinfo {volume} {118}},\ \bibinfo {pages} {131} (\bibinfo {year} {2015})}\BibitemShut {NoStop}%
\bibitem [{\citenamefont {Hall}\ and\ \citenamefont {Willitsch}(2012)}]{Felix_2012}%
  \BibitemOpen
  \bibfield  {author} {\bibinfo {author} {\bibfnamefont {F.~H.~J.}\ \bibnamefont {Hall}}\ and\ \bibinfo {author} {\bibfnamefont {S.}~\bibnamefont {Willitsch}},\ }\bibfield  {title} {\bibinfo {title} {Millikelvin reactive collisions between sympathetically cooled molecular ions and laser-cooled atoms in an ion-atom hybrid trap},\ }\href {https://doi.org/10.1103/PhysRevLett.109.233202} {\bibfield  {journal} {\bibinfo  {journal} {Phys. Rev. Lett.}\ }\textbf {\bibinfo {volume} {109}},\ \bibinfo {pages} {233202} (\bibinfo {year} {2012})}\BibitemShut {NoStop}%
\bibitem [{\citenamefont {Tomza}\ \emph {et~al.}(2019)\citenamefont {Tomza}, \citenamefont {Jachymski}, \citenamefont {Gerritsma}, \citenamefont {Negretti}, \citenamefont {Calarco}, \citenamefont {Idziaszek},\ and\ \citenamefont {Julienne}}]{Tomza_2018}%
  \BibitemOpen
  \bibfield  {author} {\bibinfo {author} {\bibfnamefont {M.}~\bibnamefont {Tomza}}, \bibinfo {author} {\bibfnamefont {K.}~\bibnamefont {Jachymski}}, \bibinfo {author} {\bibfnamefont {R.}~\bibnamefont {Gerritsma}}, \bibinfo {author} {\bibfnamefont {A.}~\bibnamefont {Negretti}}, \bibinfo {author} {\bibfnamefont {T.}~\bibnamefont {Calarco}}, \bibinfo {author} {\bibfnamefont {Z.}~\bibnamefont {Idziaszek}},\ and\ \bibinfo {author} {\bibfnamefont {P.~S.}\ \bibnamefont {Julienne}},\ }\bibfield  {title} {\bibinfo {title} {Cold hybrid ion-atom systems},\ }\href {https://doi.org/10.1103/RevModPhys.91.035001} {\bibfield  {journal} {\bibinfo  {journal} {Rev. Mod. Phys.}\ }\textbf {\bibinfo {volume} {91}},\ \bibinfo {pages} {035001} (\bibinfo {year} {2019})}\BibitemShut {NoStop}%
\bibitem [{\citenamefont {Feldker}\ \emph {et~al.}(2020)\citenamefont {Feldker}, \citenamefont {F{\"u}rst}, \citenamefont {Hirzler}, \citenamefont {Ewald}, \citenamefont {Mazzanti}, \citenamefont {Wiater}, \citenamefont {Tomza},\ and\ \citenamefont {Gerritsma}}]{Feldker_2020}%
  \BibitemOpen
  \bibfield  {author} {\bibinfo {author} {\bibfnamefont {T.}~\bibnamefont {Feldker}}, \bibinfo {author} {\bibfnamefont {H.}~\bibnamefont {F{\"u}rst}}, \bibinfo {author} {\bibfnamefont {H.}~\bibnamefont {Hirzler}}, \bibinfo {author} {\bibfnamefont {N.~V.}\ \bibnamefont {Ewald}}, \bibinfo {author} {\bibfnamefont {M.}~\bibnamefont {Mazzanti}}, \bibinfo {author} {\bibfnamefont {D.}~\bibnamefont {Wiater}}, \bibinfo {author} {\bibfnamefont {M.}~\bibnamefont {Tomza}},\ and\ \bibinfo {author} {\bibfnamefont {R.}~\bibnamefont {Gerritsma}},\ }\bibfield  {title} {\bibinfo {title} {Buffer gas cooling of a trapped ion to the quantum regime},\ }\href {https://doi.org/10.1038/s41567-019-0772-5} {\bibfield  {journal} {\bibinfo  {journal} {Nature Physics}\ }\textbf {\bibinfo {volume} {16}},\ \bibinfo {pages} {413} (\bibinfo {year} {2020})}\BibitemShut {NoStop}%
\bibitem [{\citenamefont {Drakoudis}\ \emph {et~al.}(2006)\citenamefont {Drakoudis}, \citenamefont {S{\"o}llner},\ and\ \citenamefont {Werth}}]{werthinstabilities}%
  \BibitemOpen
  \bibfield  {author} {\bibinfo {author} {\bibfnamefont {A.}~\bibnamefont {Drakoudis}}, \bibinfo {author} {\bibfnamefont {M.}~\bibnamefont {S{\"o}llner}},\ and\ \bibinfo {author} {\bibfnamefont {G.}~\bibnamefont {Werth}},\ }\bibfield  {title} {\bibinfo {title} {Instabilities of ion motion in a linear paul trap},\ }\href@noop {} {\bibfield  {journal} {\bibinfo  {journal} {International Journal of Mass Spectrometry}\ }\textbf {\bibinfo {volume} {252}},\ \bibinfo {pages} {61} (\bibinfo {year} {2006})}\BibitemShut {NoStop}%
\bibitem [{\citenamefont {Filgueira}\ \emph {et~al.}(2025)\citenamefont {Filgueira}, \citenamefont {Luda},\ and\ \citenamefont {Schmiegelow}}]{Filgueira_2025}%
  \BibitemOpen
  \bibfield  {author} {\bibinfo {author} {\bibfnamefont {L.~E.}\ \bibnamefont {Filgueira}}, \bibinfo {author} {\bibfnamefont {M.}~\bibnamefont {Luda}},\ and\ \bibinfo {author} {\bibfnamefont {C.~T.}\ \bibnamefont {Schmiegelow}},\ }\bibfield  {title} {\bibinfo {title} {Design and implementation of a blade-type linear paul trap},\ }\href {https://doi.org/10.1063/5.0301092} {\bibfield  {journal} {\bibinfo  {journal} {AIP Advances}\ }\textbf {\bibinfo {volume} {15}},\ \bibinfo {pages} {105105} (\bibinfo {year} {2025})}\BibitemShut {NoStop}%
\bibitem [{\citenamefont {Woodrow}(2015)}]{oxfordthesis}%
  \BibitemOpen
  \bibfield  {author} {\bibinfo {author} {\bibfnamefont {S.~R.}\ \bibnamefont {Woodrow}},\ }\emph {\bibinfo {title} {Linear Paul trap design for high-fidelity, scalable quantum information processing}},\ \href@noop {} {Ph.D. thesis} (\bibinfo {year} {2015})\BibitemShut {NoStop}%
\bibitem [{\citenamefont {Schwartz}\ \emph {et~al.}(2002)\citenamefont {Schwartz}, \citenamefont {Senko},\ and\ \citenamefont {Syka}}]{SCHWARTZ_2D_MS}%
  \BibitemOpen
  \bibfield  {author} {\bibinfo {author} {\bibfnamefont {J.~C.}\ \bibnamefont {Schwartz}}, \bibinfo {author} {\bibfnamefont {M.~W.}\ \bibnamefont {Senko}},\ and\ \bibinfo {author} {\bibfnamefont {J.~E.}\ \bibnamefont {Syka}},\ }\bibfield  {title} {\bibinfo {title} {A two-dimensional quadrupole ion trap mass spectrometer},\ }\href {https://doi.org/https://doi.org/10.1016/S1044-0305(02)00384-7} {\bibfield  {journal} {\bibinfo  {journal} {Journal of the American Society for Mass Spectrometry}\ }\textbf {\bibinfo {volume} {13}},\ \bibinfo {pages} {659} (\bibinfo {year} {2002})}\BibitemShut {NoStop}%
\bibitem [{\citenamefont {Joger}\ \emph {et~al.}(2017)\citenamefont {Joger}, \citenamefont {F\"urst}, \citenamefont {Ewald}, \citenamefont {Feldker}, \citenamefont {Tomza},\ and\ \citenamefont {Gerritsma}}]{Joger_2017}%
  \BibitemOpen
  \bibfield  {author} {\bibinfo {author} {\bibfnamefont {J.}~\bibnamefont {Joger}}, \bibinfo {author} {\bibfnamefont {H.}~\bibnamefont {F\"urst}}, \bibinfo {author} {\bibfnamefont {N.}~\bibnamefont {Ewald}}, \bibinfo {author} {\bibfnamefont {T.}~\bibnamefont {Feldker}}, \bibinfo {author} {\bibfnamefont {M.}~\bibnamefont {Tomza}},\ and\ \bibinfo {author} {\bibfnamefont {R.}~\bibnamefont {Gerritsma}},\ }\bibfield  {title} {\bibinfo {title} {Observation of collisions between cold li atoms and ${\mathrm{yb}}^{+}$ ions},\ }\href {https://doi.org/10.1103/PhysRevA.96.030703} {\bibfield  {journal} {\bibinfo  {journal} {Phys. Rev. A}\ }\textbf {\bibinfo {volume} {96}},\ \bibinfo {pages} {030703(R)} (\bibinfo {year} {2017})}\BibitemShut {NoStop}%
\bibitem [{\citenamefont {Meir}\ \emph {et~al.}(2018)\citenamefont {Meir}, \citenamefont {Sikorsky}, \citenamefont {Ben-shlomi}, \citenamefont {Akerman}, \citenamefont {Pinkas}, \citenamefont {Dallal},\ and\ \citenamefont {Ozeri}}]{Meir_2018}%
  \BibitemOpen
  \bibfield  {author} {\bibinfo {author} {\bibfnamefont {Z.}~\bibnamefont {Meir}}, \bibinfo {author} {\bibfnamefont {T.}~\bibnamefont {Sikorsky}}, \bibinfo {author} {\bibfnamefont {R.}~\bibnamefont {Ben-shlomi}}, \bibinfo {author} {\bibfnamefont {N.}~\bibnamefont {Akerman}}, \bibinfo {author} {\bibfnamefont {M.}~\bibnamefont {Pinkas}}, \bibinfo {author} {\bibfnamefont {Y.}~\bibnamefont {Dallal}},\ and\ \bibinfo {author} {\bibfnamefont {R.}~\bibnamefont {Ozeri}},\ }\bibfield  {title} {\bibinfo {title} {Experimental apparatus for overlapping a ground-state cooled ion with ultracold atoms},\ }\href {https://doi.org/10.1080/09500340.2017.1397217} {\bibfield  {journal} {\bibinfo  {journal} {Journal of Modern Optics}\ }\textbf {\bibinfo {volume} {65}},\ \bibinfo {pages} {501} (\bibinfo {year} {2018})},\ \Eprint {https://arxiv.org/abs/https://doi.org/10.1080/09500340.2017.1397217} {https://doi.org/10.1080/09500340.2017.1397217} \BibitemShut {NoStop}%
\bibitem [{\citenamefont {Schmid}\ \emph {et~al.}(2012)\citenamefont {Schmid}, \citenamefont {Härter}, \citenamefont {Frisch}, \citenamefont {Hoinka},\ and\ \citenamefont {Denschlag}}]{Denschlag_2012}%
  \BibitemOpen
  \bibfield  {author} {\bibinfo {author} {\bibfnamefont {S.}~\bibnamefont {Schmid}}, \bibinfo {author} {\bibfnamefont {A.}~\bibnamefont {Härter}}, \bibinfo {author} {\bibfnamefont {A.}~\bibnamefont {Frisch}}, \bibinfo {author} {\bibfnamefont {S.}~\bibnamefont {Hoinka}},\ and\ \bibinfo {author} {\bibfnamefont {J.~H.}\ \bibnamefont {Denschlag}},\ }\bibfield  {title} {\bibinfo {title} {An apparatus for immersing trapped ions into an ultracold gas of neutral atoms},\ }\href {https://doi.org/10.1063/1.4718356} {\bibfield  {journal} {\bibinfo  {journal} {Review of Scientific Instruments}\ }\textbf {\bibinfo {volume} {83}},\ \bibinfo {pages} {053108} (\bibinfo {year} {2012})},\ \url{https://pubs.aip.org/aip/rsi/article-pdf/doi/10.1063/1.4718356/13546646/053108_1_online.pdf} \BibitemShut {NoStop}%
\bibitem [{\citenamefont {Heinrich}(2018)}]{DD_Heinrich2018BeTrap}%
  \BibitemOpen
  \bibfield  {author} {\bibinfo {author} {\bibfnamefont {J.~M.}\ \bibnamefont {Heinrich}},\ }{\emph {\bibinfo {title} {{A Be$^{+}$ Ion Trap for H$_2^{+}$ Spectroscopy}}}},\ \href {https://theses.hal.science/tel-02049379} {\bibinfo {type} {Ph.d. thesis}},\ \bibinfo  {school} {Sorbonne Université, Faculté des Sciences et Ingénierie} (\bibinfo {year} {2018}),\ \bibinfo {note} {hAL Thesis: tel-02049379v1}\BibitemShut {NoStop}%
\bibitem [{\citenamefont {Hempel}(2014)}]{DD_Hempel2014Thesis}%
  \BibitemOpen
  \bibfield  {author} {\bibinfo {author} {\bibfnamefont {C.}~\bibnamefont {Hempel}},\ }\emph {\bibinfo {title} {Digital Quantum Simulation, Schr{\"o}dinger Cat State Spectroscopy and Setting Up a Linear Ion Trap}},\ \href@noop {} {\bibinfo {type} {Doctoral thesis}} (\bibinfo {year} {2014}),\ \bibinfo {note} {institute for Quantum Optics and Quantum Information (IQOQI), Austrian Academy of Sciences}\BibitemShut {NoStop}%
\bibitem [{\citenamefont {Ballance}(2014)}]{DD_Ballance2014}%
  \BibitemOpen
  \bibfield  {author} {\bibinfo {author} {\bibfnamefont {C.~J.}\ \bibnamefont {Ballance}},\ }\href@noop {} {\bibinfo {title} {High-fidelity quantum logic in {Ca}$^{+}$}} (\bibinfo {year} {2014}),\ \bibinfo {note} {department of Physics, Hertford College}\BibitemShut {NoStop}%
\bibitem [{\citenamefont {Dahl}(2022)}]{simion}%
  \BibitemOpen
  \bibfield  {author} {\bibinfo {author} {\bibfnamefont {D.~A.}\ \bibnamefont {Dahl}},\ }\href {https://simion.com} {\bibinfo {title} {Simion 8.x}} (\bibinfo {year} {2022})\BibitemShut {NoStop}%
\bibitem [{\citenamefont {{Wolfram Research, Inc.}}(2023)}]{mathematica}%
  \BibitemOpen
  \bibfield  {author} {\bibinfo {author} {\bibnamefont {{Wolfram Research, Inc.}}},\ }\href {https://www.wolfram.com/mathematica} {\bibinfo {title} {Mathematica, version 13.x}} (\bibinfo {year} {2023})\BibitemShut {NoStop}%
\bibitem [{\citenamefont {Joshi}\ \emph {et~al.}(2026)\citenamefont {Joshi}, \citenamefont {Mahendrakar}, \citenamefont {Niranjan}, \citenamefont {Yadav}, \citenamefont {Krishnakumar}, \citenamefont {Pandey}, \citenamefont {Vexiau}, \citenamefont {Dulieu},\ and\ \citenamefont {Rangwala}}]{joshi_AI}%
  \BibitemOpen
  \bibfield  {author} {\bibinfo {author} {\bibfnamefont {N.}~\bibnamefont {Joshi}}, \bibinfo {author} {\bibfnamefont {V.}~\bibnamefont {Mahendrakar}}, \bibinfo {author} {\bibfnamefont {M.}~\bibnamefont {Niranjan}}, \bibinfo {author} {\bibfnamefont {R.~S.}\ \bibnamefont {Yadav}}, \bibinfo {author} {\bibfnamefont {E.}~\bibnamefont {Krishnakumar}}, \bibinfo {author} {\bibfnamefont {A.}~\bibnamefont {Pandey}}, \bibinfo {author} {\bibfnamefont {R.}~\bibnamefont {Vexiau}}, \bibinfo {author} {\bibfnamefont {O.}~\bibnamefont {Dulieu}},\ and\ \bibinfo {author} {\bibfnamefont {S.~A.}\ \bibnamefont {Rangwala}},\ }\bibfield  {title} {\bibinfo {title} {Efficient associative ionization of ultracold lithium and the long-lived ${\mathrm{li}}_{2}^{+}$},\ }\href {https://doi.org/10.1103/z7xy-hmlb} {\bibfield  {journal} {\bibinfo  {journal} {Phys. Rev. Res.}\ }\textbf {\bibinfo {volume} {8}},\ \bibinfo {pages} {L032008} (\bibinfo {year} {2026})}\BibitemShut {NoStop}%
\bibitem [{jos()}]{joshi_Instrumentation}%
  \BibitemOpen
  \bibfield  {title} {\bibinfo {title} {Details about the instrument will be reported in a separate publication.},\ }\href@noop {} {\ }\BibitemShut {NoStop}%
\bibitem [{\citenamefont {Zhao}\ \emph {et~al.}(2002)\citenamefont {Zhao}, \citenamefont {Ryjkov},\ and\ \citenamefont {Schuessler}}]{PRA_Zhao_2002}%
  \BibitemOpen
  \bibfield  {author} {\bibinfo {author} {\bibfnamefont {X.}~\bibnamefont {Zhao}}, \bibinfo {author} {\bibfnamefont {V.~L.}\ \bibnamefont {Ryjkov}},\ and\ \bibinfo {author} {\bibfnamefont {H.~A.}\ \bibnamefont {Schuessler}},\ }\bibfield  {title} {\bibinfo {title} {Parametric excitations of trapped ions in a linear rf ion trap},\ }\href {https://doi.org/10.1103/PhysRevA.66.063414} {\bibfield  {journal} {\bibinfo  {journal} {Phys. Rev. A}\ }\textbf {\bibinfo {volume} {66}},\ \bibinfo {pages} {063414} (\bibinfo {year} {2002})}\BibitemShut {NoStop}%
\bibitem [{\citenamefont {Schmidt}\ \emph {et~al.}(2020{\natexlab{a}})\citenamefont {Schmidt}, \citenamefont {H\"onig}, \citenamefont {Weckesser}, \citenamefont {Thielemann}, \citenamefont {Schaetz},\ and\ \citenamefont {Karpa}}]{APB_Schmidt_2020}%
  \BibitemOpen
  \bibfield  {author} {\bibinfo {author} {\bibfnamefont {J.}~\bibnamefont {Schmidt}}, \bibinfo {author} {\bibfnamefont {D.}~\bibnamefont {H\"onig}}, \bibinfo {author} {\bibfnamefont {P.}~\bibnamefont {Weckesser}}, \bibinfo {author} {\bibfnamefont {F.}~\bibnamefont {Thielemann}}, \bibinfo {author} {\bibfnamefont {T.}~\bibnamefont {Schaetz}},\ and\ \bibinfo {author} {\bibfnamefont {L.}~\bibnamefont {Karpa}},\ }\bibfield  {title} {\bibinfo {title} {Mass-selective removal of ions from paul traps using parametric excitation},\ }\href {https://doi.org/10.1007/s00340-020-07491-8} {\bibfield  {journal} {\bibinfo  {journal} {Appl. Phys. B}\ }\textbf {\bibinfo {volume} {126}} (\bibinfo {year} {2020}{\natexlab{a}})}\BibitemShut {NoStop}%
\bibitem [{\citenamefont {Jyothi}\ \emph {et~al.}(2019)\citenamefont {Jyothi}, \citenamefont {Egodapitiya}, \citenamefont {Bondurant}, \citenamefont {Jia}, \citenamefont {Pretzsch}, \citenamefont {Chiappina}, \citenamefont {Shu},\ and\ \citenamefont {Brown}}]{Jyothi_2019}%
  \BibitemOpen
  \bibfield  {author} {\bibinfo {author} {\bibfnamefont {S.}~\bibnamefont {Jyothi}}, \bibinfo {author} {\bibfnamefont {K.~N.}\ \bibnamefont {Egodapitiya}}, \bibinfo {author} {\bibfnamefont {B.}~\bibnamefont {Bondurant}}, \bibinfo {author} {\bibfnamefont {Z.}~\bibnamefont {Jia}}, \bibinfo {author} {\bibfnamefont {E.}~\bibnamefont {Pretzsch}}, \bibinfo {author} {\bibfnamefont {P.}~\bibnamefont {Chiappina}}, \bibinfo {author} {\bibfnamefont {G.}~\bibnamefont {Shu}},\ and\ \bibinfo {author} {\bibfnamefont {K.~R.}\ \bibnamefont {Brown}},\ }\bibfield  {title} {\bibinfo {title} {A hybrid ion-atom trap with integrated high resolution mass spectrometer},\ }\href {https://doi.org/10.1063/1.5121431} {\bibfield  {journal} {\bibinfo  {journal} {Review of Scientific Instruments}\ }\textbf {\bibinfo {volume} {90}},\ \bibinfo {pages} {103201} (\bibinfo {year} {2019})},\ \url {https://arxiv.org/abs/https://pubs.aip.org/aip/rsi/article-pdf/doi/10.1063/1.5121431/19755577/103201_1_online.pdf}
 \BibitemShut {NoStop}%
\bibitem [{\citenamefont {Schmidt}\ \emph {et~al.}(2020{\natexlab{b}})\citenamefont {Schmidt}, \citenamefont {Weckesser}, \citenamefont {Thielemann}, \citenamefont {Schaetz},\ and\ \citenamefont {Karpa}}]{Schmidt_2020}%
  \BibitemOpen
  \bibfield  {author} {\bibinfo {author} {\bibfnamefont {J.}~\bibnamefont {Schmidt}}, \bibinfo {author} {\bibfnamefont {P.}~\bibnamefont {Weckesser}}, \bibinfo {author} {\bibfnamefont {F.}~\bibnamefont {Thielemann}}, \bibinfo {author} {\bibfnamefont {T.}~\bibnamefont {Schaetz}},\ and\ \bibinfo {author} {\bibfnamefont {L.}~\bibnamefont {Karpa}},\ }\bibfield  {title} {\bibinfo {title} {Optical traps for sympathetic cooling of ions with ultracold neutral atoms},\ }\href {https://doi.org/10.1103/PhysRevLett.124.053402} {\bibfield  {journal} {\bibinfo  {journal} {Phys. Rev. Lett.}\ }\textbf {\bibinfo {volume} {124}},\ \bibinfo {pages} {053402} (\bibinfo {year} {2020}{\natexlab{b}})}\BibitemShut {NoStop}%
\bibitem [{\citenamefont {Ballance}\ \emph {et~al.}(2016)\citenamefont {Ballance}, \citenamefont {Harty}, \citenamefont {Linke}, \citenamefont {Sepiol},\ and\ \citenamefont {Lucas}}]{DD_Used_CJBalance_PRL}%
  \BibitemOpen
  \bibfield  {author} {\bibinfo {author} {\bibfnamefont {C.~J.}\ \bibnamefont {Ballance}}, \bibinfo {author} {\bibfnamefont {T.~P.}\ \bibnamefont {Harty}}, \bibinfo {author} {\bibfnamefont {N.~M.}\ \bibnamefont {Linke}}, \bibinfo {author} {\bibfnamefont {M.~A.}\ \bibnamefont {Sepiol}},\ and\ \bibinfo {author} {\bibfnamefont {D.~M.}\ \bibnamefont {Lucas}},\ }\bibfield  {title} {\bibinfo {title} {High-fidelity quantum logic gates using trapped-ion hyperfine qubits},\ }\href {https://doi.org/10.1103/PhysRevLett.117.060504} {\bibfield  {journal} {\bibinfo  {journal} {Phys. Rev. Lett.}\ }\textbf {\bibinfo {volume} {117}},\ \bibinfo {pages} {060504} (\bibinfo {year} {2016})}\BibitemShut {NoStop}%
\bibitem [{\citenamefont {Joshi}\ \emph {et~al.}(2020)\citenamefont {Joshi}, \citenamefont {Fabre}, \citenamefont {Maier}, \citenamefont {Brydges}, \citenamefont {Kiesenhofer}, \citenamefont {Hainzer}, \citenamefont {Blatt},\ and\ \citenamefont {Roos}}]{DD_used_Joshi2020PGC}%
  \BibitemOpen
  \bibfield  {author} {\bibinfo {author} {\bibfnamefont {M.~K.}\ \bibnamefont {Joshi}}, \bibinfo {author} {\bibfnamefont {A.}~\bibnamefont {Fabre}}, \bibinfo {author} {\bibfnamefont {C.}~\bibnamefont {Maier}}, \bibinfo {author} {\bibfnamefont {T.}~\bibnamefont {Brydges}}, \bibinfo {author} {\bibfnamefont {D.}~\bibnamefont {Kiesenhofer}}, \bibinfo {author} {\bibfnamefont {H.}~\bibnamefont {Hainzer}}, \bibinfo {author} {\bibfnamefont {R.}~\bibnamefont {Blatt}},\ and\ \bibinfo {author} {\bibfnamefont {C.~F.}\ \bibnamefont {Roos}},\ }\bibfield  {title} {\bibinfo {title} {Polarization-gradient cooling of {1D} and {2D} ion coulomb crystals},\ }\href {https://doi.org/10.1088/1367-2630/abb912} {\bibfield  {journal} {\bibinfo  {journal} {New Journal of Physics}\ }\textbf {\bibinfo {volume} {22}},\ \bibinfo {pages} {103013} (\bibinfo {year} {2020})}\BibitemShut {NoStop}%
\bibitem [{\citenamefont {Kaewuam}\ \emph {et~al.}(2018)\citenamefont {Kaewuam}, \citenamefont {Roy}, \citenamefont {Tan}, \citenamefont {Arnold},\ and\ \citenamefont {Barrett}}]{DD_used_Kaewuam_2018}%
  \BibitemOpen
  \bibfield  {author} {\bibinfo {author} {\bibfnamefont {R.}~\bibnamefont {Kaewuam}}, \bibinfo {author} {\bibfnamefont {A.}~\bibnamefont {Roy}}, \bibinfo {author} {\bibfnamefont {T.~R.}\ \bibnamefont {Tan}}, \bibinfo {author} {\bibfnamefont {K.~J.}\ \bibnamefont {Arnold}},\ and\ \bibinfo {author} {\bibfnamefont {M.~D.}\ \bibnamefont {Barrett}},\ }\bibfield  {title} {\bibinfo {title} {Laser spectroscopy of 176lu+},\ }\href {https://doi.org/10.1080/09500340.2017.1411539} {\bibfield  {journal} {\bibinfo  {journal} {Journal of Modern Optics}\ }\textbf {\bibinfo {volume} {65}},\ \bibinfo {pages} {592} (\bibinfo {year} {2018})},\ \Eprint {https://arxiv.org/abs/https://doi.org/10.1080/09500340.2017.1411539} {https://doi.org/10.1080/09500340.2017.1411539} \BibitemShut {NoStop}%
\bibitem [{\citenamefont {Olmschenk}\ \emph {et~al.}(2009)\citenamefont {Olmschenk}, \citenamefont {Matsukevich}, \citenamefont {Maunz}, \citenamefont {Hayes}, \citenamefont {Duan},\ and\ \citenamefont {Monroe}}]{DD_used_Olmschenk}%
  \BibitemOpen
  \bibfield  {author} {\bibinfo {author} {\bibfnamefont {S.}~\bibnamefont {Olmschenk}}, \bibinfo {author} {\bibfnamefont {D.~N.}\ \bibnamefont {Matsukevich}}, \bibinfo {author} {\bibfnamefont {P.}~\bibnamefont {Maunz}}, \bibinfo {author} {\bibfnamefont {D.}~\bibnamefont {Hayes}}, \bibinfo {author} {\bibfnamefont {L.-M.}\ \bibnamefont {Duan}},\ and\ \bibinfo {author} {\bibfnamefont {C.}~\bibnamefont {Monroe}},\ }\bibfield  {title} {\bibinfo {title} {Quantum teleportation between distant matter qubits},\ }\href {https://doi.org/10.1126/science.1167209} {\bibfield  {journal} {\bibinfo  {journal} {Science}\ }\textbf {\bibinfo {volume} {323}},\ \bibinfo {pages} {486} (\bibinfo {year} {2009})}\BibitemShut {NoStop}%
\bibitem [{\citenamefont {Surendra}\ \emph {et~al.}(2026)\citenamefont {Surendra}, \citenamefont {Hoffmann},\ and\ \citenamefont {Köhl}}]{Surendra_2026}%
  \BibitemOpen
  \bibfield  {author} {\bibinfo {author} {\bibfnamefont {S.}~\bibnamefont {Surendra}}, \bibinfo {author} {\bibfnamefont {A.}~\bibnamefont {Hoffmann}},\ and\ \bibinfo {author} {\bibfnamefont {M.}~\bibnamefont {Köhl}},\ }\href {https://doi.org/10.48550/arXiv.2602.07700} {\bibinfo {title} {Two-phase driving of a linear radio-frequency ion trap}} (\bibinfo {year} {2026}),\ \Eprint {https://arxiv.org/abs/2602.07700} {arXiv:2602.07700 [quant-ph]} \BibitemShut {NoStop}%
\end{thebibliography}
%

\end{document}